\documentclass[a4paper]{article}
\usepackage{epsfig}
\begin{document}

\title{\bf\large Measurement of the decay  $\phi\to\mu^+\mu^-$.}

\author{
M.N.Achasov, 
S.E.Baru,
A.V.Bozhenok, 
D.A.Bukin, 
\underline{S.V.Burdin\thanks{E-mail: burdin@inp.nsk.su}},
 \\
T.V.Dimova, 
V.P.Druzhinin, 
M.S.Dubrovin, 
I.A.Gaponenko, \\ 
V.B.Golubev,
V.N.Ivanchenko, 
P.I.Ivanov, \\
A.A.Korol, 
S.V.Koshuba,
E.A.Perevedentsev, 
E.E.Pyata, \\
A.A.Salnikov, 
S.I.Serednyakov,
V.V.Shary, 
Yu.M.Shatunov, \\ 
V.A.Sidorov, 
Z.K.Silagadze,
Yu.V.Usov
\\ \\
{\it Budker Institute of Nuclear Physics,} \\ 
{\it 630090, Novosibirsk, } \\ 
{\it Russia }
}

\date{}
\maketitle

\begin{abstract}
\normalsize
The process $e^+e^-\to\mu^+\mu^-$ 
has been studied with SND detector at VEPP-2M
$e^+e^-$ collider in the vicinity of  $\phi(1020)$
resonance. 
The product of branching ratios
of $\phi$ meson into leptons \\
$\sqrt{B(\phi\to\mu^+\mu^-)\cdot B(\phi\to
e^+e^-)}=(3.14\pm0.22\pm0.14)\cdot 10^{-4}$ 
was measured from the interference in the cross section of the
process $e^+e^-\to\mu^+\mu^-$.
The branching ratio 
$B(\phi\to\mu^+\mu^-)=(3.30\pm0.45\pm0.32)\cdot 10^{-4}$ was obtained. 
\end{abstract}

\twocolumn

\section{Introduction}
\normalsize
The $\phi\to\mu^+\mu^-$ decay reveals itself as an
interference pattern in the energy
dependence of the cross section of process $e^+e^-\to\mu^+\mu^-$ 
in the region around $\phi$-resonance peak.
The interference amplitude is determined by the branching ratio
of the decay $\phi\to\mu^+\mu^-$.
The table value of the branching ratio
$B(\phi\to\mu^+\mu^-)=(2.5\pm0.4)\cdot 10^{-4}$ \cite{PDG} 
is based on the experiments on photoproduction of $\phi$ meson 
\cite{Earles,Hayes}.
In $e^+e^-$ collisions directly measurable is the square root of
the product of the branching ratios \\
$\sqrt{B(\phi\to\mu^+\mu^-)\cdot B(\phi\to e^+e^-)}$.
The $\mu-e$-universality requires  $B(\phi\to\mu^+\mu^-)$ to be equal
to $B(\phi\to e^+e^-)$ with the accuracy much higher than our
experimental one despite the difference in  masses of the leptons.
The measurements of the branching ratio $B(\phi\to e^+e^-)$ 
at $e^+e^-$ colliders
are performed through
summation of cross sections over all decay channels \cite{CMD2a}. 
The study of the interference
$e^+e^-\to\phi\to\mu^+\mu^-$
 gives the independent evaluation 
of the $\phi$-meson leptonic width.
First measurement of the decay $\phi\to\mu^+\mu^-$ on $e^+e^-$
collider 
was performed in Orsay in 1972 \cite{ORSAY}. In this experiment the 
 value of branching ratio of $\phi$-meson leptonic decay \\
 $\sqrt{B(\phi\to\mu^+\mu^-)\cdot B(\phi\to e^+e^-)}=
(2.93\pm0.96\pm0.32)\cdot 10^{-4}$ was
obtained. Later similar measurements were
performed in Novosibirsk \cite{Fipipi,CMDprep}.

\section{ Experiment}
\begin{figure}[htb]
\begin{minipage}[t]{0.475\textwidth}
\centerline{\includegraphics[width=8cm]{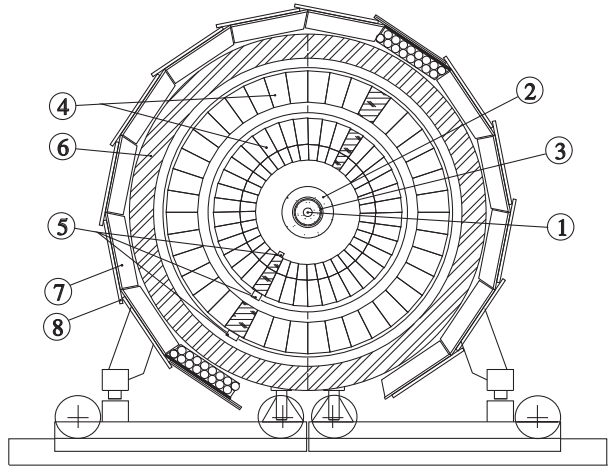}}
\caption{
Detector  SND  --- view across the beam;
1 --- beam pipe, 2 --- drift chambers, 3 --- inner scintillation
counters, 4 
--- NaI(Tl) counters, 5 --- vacuum phototriodes, 6 --- iron absorber, 7 
--- streamer tubes, 8 --- outer scintillation counters
}
\label{SNDtrans}
\end{minipage}
\end{figure}

\normalsize
The experiment was carried out with SND detector (Fig.~\ref{SNDtrans})
at VEPP-2M in 1996--1997. SND is a general purpose non-magnetic
detector \cite{SND,Prep.97}. The main part of the SND is a spherical
electromagnetic calorimeter, consisting of 1630 NaI(Tl) crystals. The
solid angle of the calorimeter is $\sim 90\%$ of $4\pi$
steradian. The angles of charged particles  are
measured by two cylindrical drift chambers covering 95\% of $4\pi$
steradian solid angle. The important part of the detector for the
process under study is a muon system, consisting of streamer tubes
and plastic scintillation counters, with
 $1$~cm iron plates between blocks of tubes and counters. 
 The simultaneous hits in the streamer tubes and scintillation counters
produce a signal of muon system.
The time difference between the hit and
the beam collision is measured by inner \cite{BukinD} and outer
scintillation counters \cite{Bagdat}. 

  The experiment was carried out in the energy range
  $2E_b=984$--1040 MeV and consisted of 6 data taking runs 
\cite{Prep.98,Had97}: \\ PHI\_9601 -- PHI\_9606. 
Five runs were used for analysis, corresponding to the total
  integrated luminosity  
$\Delta L=2.61$~pb$^{-1}$. \\ About $4.6\cdot 10^{6}$ of $\phi$ mesons were
 produced.
The integrated luminosity was measured with the accuracy of about 3\%
using $e^+e^-\to e^+e^-$ and $e^+e^-\to\gamma\gamma$  events.

\section{Event selection} 
\normalsize
  The energy behavior of the cross section of the process 
\begin{equation}
 e^{+}e^{-} \to\mu^+\mu^-(\gamma)
\label{mumu}
\end{equation}
was studied in the vicinity of $\phi$ meson. Events with two
collinear charged particles were selected for analysis. 
The selection of the events of the process
(\ref{mumu}) was performed with the following cuts on  angles of acollinearity
 of the charged particles in azimuth and polar directions:
$\mid\Delta\phi\mid<10^\circ$, $\mid\Delta\theta\mid<25^\circ$. 
Additional
photon emitted
by either initial or final particles was permitted.
To avoid possible
 losses of events due to beam background or
knock-on electrons in the drift chambers, 
one additional charged particle was also permitted. 
To suppress the beam background the production point of charged
particles was required to be within $0.5$~cm from the 
interaction point in the azimuth plane and $\pm 7.5$~cm along 
the beam direction (the longitudinal size of the interaction region 
 $\sigma_{z}$ is about $2$~cm). 
The polar angles of the charged particles were limited to the range 
$45^\circ<\theta<135^\circ$, corresponding to the acceptance angle of 
the muon system.

The main sources of background are the cosmic muons and the processes
\begin{equation}
 e^{+}e^{-} \to e^{+}e^{-},
\label{ee}
\end{equation} 
\begin{equation}
 e^{+}e^{-} \to \pi^{+}\pi^{-}(\gamma).
\label{pipi}
\end{equation}
To suppress the background from the process (\ref{ee}) a procedure of
$e/\pi(\mu)$ separation was used. 
The algorithm is similar to that, developed for the
 ND detector \cite{NDepi}. It utilizes the difference in total energy
depositions and 
the longitudinal energy deposition profiles
for electrons, pions and muons.
As a result of this procedure the background from the process
(\ref{ee}) was suppressed down to 4\% of the events of the
process (\ref{mumu}). To suppress the contribution of the process
(\ref{pipi}) a procedure of $\pi/\mu$ separation by the muon system has
been used. In the energy range under study the probability to hit the muon
system for a muon varies from 80 to 93\% and is as low as 1.5\% for
pions of the process (\ref{pipi}). 
A requirement of a hit in the
muon system also reduced the background from the process (\ref{ee}) by 
two order of magnitude, making its contribution negligible.

After these cuts 80\% of selected events are still  
 cosmic background. The
rejection of the events with the hit in two top segments of the muon system  
(about $45^\circ$ in azimuth direction)
(Fig.~\ref{SNDtrans}) suppressed the cosmic events by two times. 
Further suppression of the cosmic background was performed using
the following parameters:
\begin{enumerate}
\item 
the time difference between the hit of the muon system and 
 the beam collision -- $\tau$;
\item
the time difference between the hits in upper and lower halfs of the muon
system -- $TOF$;
\item
the sum of the distances from the tracks to the production point -- $R$; 
\item
the likelihood function -- $P_\mu$, which is built on the basis of the
energy depositions in the calorimeter layers for the muon: 
\begin{equation}
 P_\mu=P_{\mu 1}(E_1)\cdot P_{\mu 2}(E_2)\cdot P_{\mu 3}(E_3),
\label{Pmu}
\end{equation}
where $P_{\mu i}(E_i)$ is the value of the probability density function
for the energy depositions $E_i$ in $i$-th calorimeter layer. These functions
were obtained from the data sample, where muons were selected by muon system
using strict cuts ($\mid\tau\mid<5$~ns, $TOF<0$~ns, see explanation below).
\end{enumerate}

\begin{figure}
\begin{minipage}[t]{0.475\textwidth}
\epsfig{figure=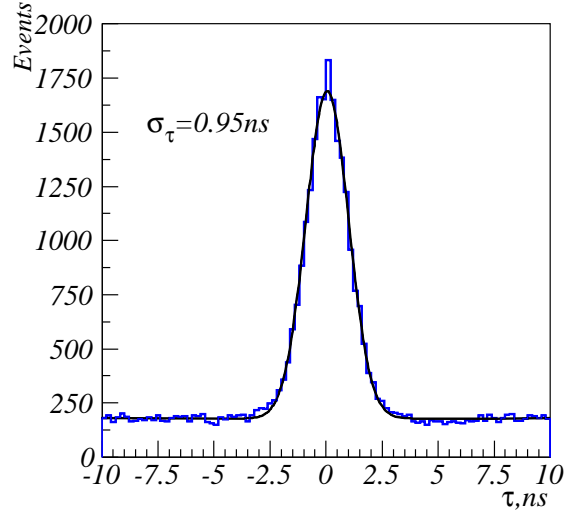}
\large
\caption{
The time of the particle passage through the upper half of the muon
system with respect to the beam collision time.}
\label{timeup}
\end{minipage}
\end{figure}
\begin{figure}
\epsfig{figure=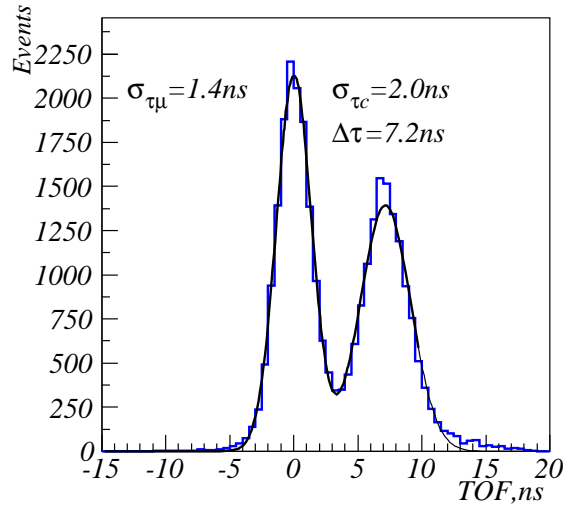}
\large
\caption{
The time of flight $TOF$ of the particle from the upper half of the muon
system to the lower one. The peak at 0 is due to events of the
$e^+e^-$ interaction, the right peak --- cosmic background.}
\label{timefly}
\end{figure}
\begin{figure}
\epsfig{figure=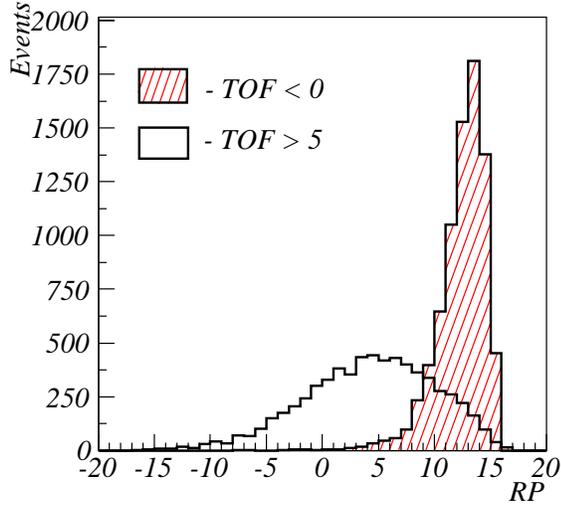}
\large
\caption{
The $RP$ distribution for
the events with $TOF<0$~ns
(hatched histogram) and
events with $TOF>5$~ns (empty histogram).}
\label{promu}
\end{figure}
The distribution of the parameter $\tau$
is shown in Fig. \ref{timeup}.
The scale for parameter $\tau$ was adjusted to have a peak at $0$ for
$e^+e^- \to \mu^+\mu^-$ events.
The events with $\mid\tau\mid<5$~ns were selected for further
analysis. 
In about 75\% of the selected events there is a hit in the lower half
of the muon system.
For these events Fig. \ref{timefly} shows the distribution of 
the time of flight
$TOF$. Peaks in the spectra at $TOF=0$~ns and $TOF=7.2$~ns are
due to the process (\ref{mumu}) and cosmic background
respectively.  

To subtract the cosmic background the combination of the independent
parameters $R$ and $P_\mu$ was used: 
\begin{equation}
 RP=\ln(P_{\mu})-16 \cdot R(cm) +25.
\label{RP}
\end{equation}   
The factor before $R$ has been chosen to provide maximum separation between
cosmic events and the events of the process (\ref{mumu}).
Additive constant equal to $25$ fixes the scale of the parameter $RP$. 
Fig. \ref{promu} shows the $RP$ distribution 
for the events with $TOF>5$~ns (mainly cosmic events) and for the events
with $TOF<0$~ns (mainly events of the process (\ref{mumu})).
Most of the events of the process (\ref{mumu}) are in the region  $RP>0$. 
At each energy point
the number of events of the process (\ref{mumu}) was estimated by the
following formula:
\begin{equation}
  N_\mu=N_{RP>0}-C\cdot N_{RP<0}.
\label{Nmu}
\end{equation}
Here $C=\frac{N_{RP>0}'}{N_{RP<0}'}$; 
$N_{RP>0}'$ and $N_{RP<0}'$ were
determined from the sample of the events with $TOF>5$~ns; $N_{RP>0}$,
$N_{RP<0}$ --- from the sample of the events with $TOF<5$~ns. 

 The detection efficiency $\varepsilon_\mu$ for the process
 (\ref{mumu}) was determined from the Monte Carlo simulation.
 The events generator was based on the formulae from the work \cite{theor}.
 The passage of the particles through the detector
 was simulated by the program UNIMOD2
 \cite{UNIMOD}. The energy dependence of the detection efficiency
 is  especially important for the process under study. This dependence
 is determined mainly by the probability for a muon to reach the muon system.  

 The detection efficiency determined from the Monte Carlo was corrected
 in order to take into account the effects, which were
 not included into simulation. First  
 is the efficiency of a software 3-rd level trigger, which selected the
 events during the data taking. This efficiency
 equals to 97--98\% for all runs, except the run PHI\_9601, where it was
 93\%. Second effect not included into the simulation is 
 the inefficiency of the muon system. It was determined from the
experimental data for each energy point and was also used as a
 correction for overall efficiency.
 This correction is about $9\%$ and was determined mainly by
 the dead time of the time channels.

\section{Data analysis}
\normalsize

The energy dependence of
the detection cross section was fitted with the
following formula:
\begin{equation}
 \sigma_{vis}= \varepsilon_\mu(E)\sigma_{\mu\mu(\gamma)}(E)+
 \varepsilon_\pi\sigma_{\pi\pi(\gamma)}(E),
\label{crossapp}
\end{equation}
where $E=2E_b$;  $\varepsilon_\mu(E)$ 
and $\varepsilon_\pi\approx 0.017$ --- the detection efficiencies of
the processes (\ref{mumu}) and (\ref{pipi}), $\sigma_{\mu\mu(\gamma)}$
and $\sigma_{\pi\pi(\gamma)}$ --- the cross sections of these
processes. The energy dependence of the efficiency obtained from the
simulation was approximated by a smooth function. 

To perform a combined fit of the detection cross sections 
for all experimental runs the scale factors $\varepsilon_{sf}^i$ 
 for each run were introduced into the
detection efficiencies as free fit parameters. 
They are asymptotic detection efficiencies at 
energies higher than 520 MeV, because the probability for a muon with
such energy to hit the muon system is almost constant. This method
provides an evaluation of efficiencies independent from simulation.

The cross section
$\sigma_{\pi\pi(\gamma)}$ was calculated according to formulae
from the work \cite{theor} and taking into account the experimental
data on the pion form factor \cite{formpi} and the decay $\phi\to 2\pi$
\cite{PDG}. The cross section $\sigma_{\mu\mu(\gamma)}(E)$ was taken
in the following form:
\begin{equation}
\sigma_{\mu\mu(\gamma)}(E) = \sigma_{\mu\mu}(E)\cdot \beta(E),
\label{fitcross2}
\end{equation}
$$\sigma_{\mu\mu}(E)=83.50(nb)\frac{m_{\phi}^2}{E^2} \mid Z \mid ^2,$$
$$Z = 1 - Q\cdot e^{i\psi_\mu}\cdot \frac{m_{\phi}\Gamma_{\phi}}
{m_{\phi}^{2} - E^2 - iE\Gamma(E)},$$
where
$\sigma_{\mu\mu}(E)$ is the Born
 cross section of the process $e^+e^-\to\mu^+\mu^- $;
$m_{\phi},\Gamma_{\phi}$
--- mass and width of $\phi$ meson;
 $Q,\psi_\mu$ --- module and phase of the interference amplitude;
$\beta(E)$ --- factor taking into account the radiative corrections.
This factor was obtained as a ratio of the cross section of process
(\ref{mumu}) to the Born cross section for the same acceptance
angles and with the $\phi$-meson contribution. The cross section of the
process (\ref{mumu}) was calculated 
by the Monte Carlo method using the formulae from the work
\cite{theor} with appropriate 
cuts on the angles and momenta of the final muons.

The luminosity measurement was performed taking into account 
the interference term in the cross section of the process (\ref{ee}).
 The interference amplitude is about $0.5\%$ and the phase
differs by 180 degrees from the phase in the
process (\ref{mumu}).

The fit of the experimental data gives the interference phase
$\psi_\mu=(4.5\pm3.4)^\circ$, which is consistent with the
theoretical expectation of  $0^\circ$. Therefore 
the final fit was made with the fixed phase 
$\psi_\mu=0^\circ$. As a result the value of the interference amplitude
$$Q=0.129\pm 0.009$$ and the
 detection efficiencies $\varepsilon_{sf}^i$ for five runs were
obtained (table \ref{tabef}). 
\begin{table}[htb]
\large
\caption
{
The detection efficiencies $\varepsilon_{sf}^i$ for the process
(\ref{mumu}), 
obtained from the simulation and the experiment.
}
\label{tabef}
\begin{tabular}{ccc}
\hline
No & $\varepsilon_{sf}^i,MC$   &  $\varepsilon_{sf}^i,exp$      \\
\hline
PHI\_9601 & 0.270$\pm$0.003 & 0.254$\pm$0.005 \\
PHI\_9602 & 0.282$\pm$0.004 & 0.277$\pm$0.004 \\
PHI\_9604 & 0.287$\pm$0.003 & 0.292$\pm$0.003 \\
PHI\_9605 & 0.289$\pm$0.003 & 0.290$\pm$0.003 \\
PHI\_9606 & 0.287$\pm$0.003 & 0.260$\pm$0.003 \\
\hline
\end{tabular}
\end{table}

One can see from the table that the detection efficiencies,
obtained from the simulation and the experiment, are in good
agreement, except for the runs PHI\_9601 and PHI\_9606. 
In PHI\_9601 run the muon system worked in different
 regime. It was not fully compensated by the corrections. 
In addition, the efficiency of the software trigger in this run was
lower. The
explanation of the lower efficiency in the run PHI\_9606 is a
 lower gain in the drift chambers and consequently
 a worse resolution in $\theta$.

The error in the detection efficiency does not directly contribute
into the error at interference amplitude if each run is
recorded in the same conditions
because the interference amplitude is a relative value.  
Therefore the data with low
gain in the drift chambers were excluded from the runs.

\begin{figure*}
\begin{minipage}[htb]{0.99\textwidth}
\centerline{\includegraphics[width=16cm,height=7cm]{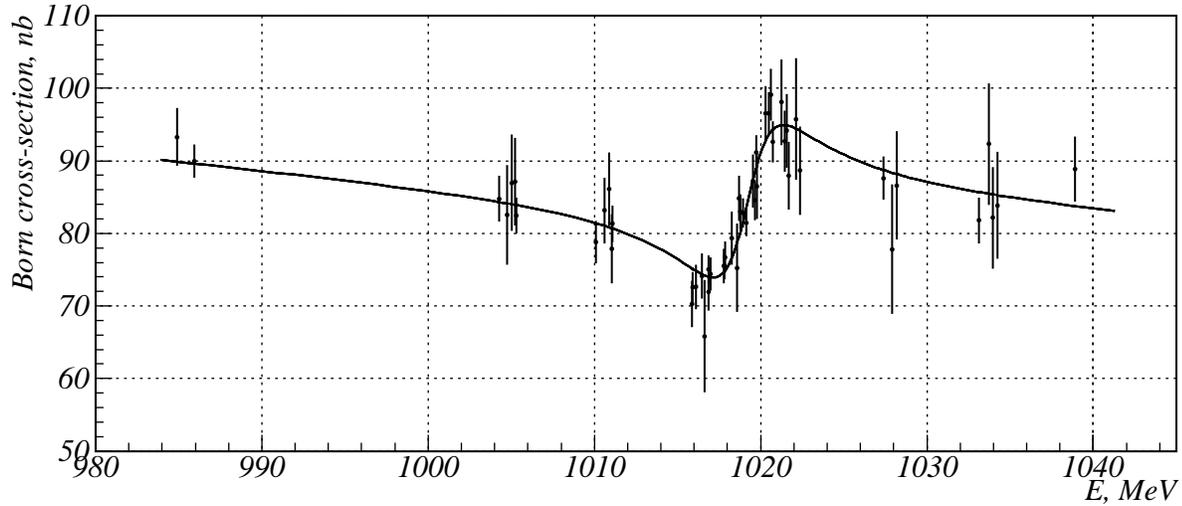}}
\caption{
The energy dependence of Born cross section of the process
$e^+e^-\to\mu^+\mu^-$ with $\phi$-meson contribution.
}
\label{secall}
\end{minipage}
\end{figure*}
Fig. \ref{secall} shows the Born cross section of the process
$e^+e^-\to\mu^+\mu^-$. The $\chi^2$ value 
 obtained in the fit equals to $36.0$ for 45 degrees of
freedom.

The systematic error of the amplitude of the interference is
determined by the errors of the luminosity estimation and the
calculation of the radiative corrections. The measurement of the
luminosity using the events of the process
$e^+e^-\to\gamma\gamma$ gives the estimation of the systematic error
$\delta_{lum}=2.1\%$.
Additional systematic error can be due to imprecise
evaluation of the background from the process (\ref{pipi}).
The estimation of this contribution is $\delta_{\pi\pi}=0.6\%$.
The contribution of the error in the calculation of the 
radiative corrections to the systematic errors of the 
interference amplitude is $\delta_\beta=4\%$. 
The resulting systematic error is $\delta=4.6\%$.  

The interference amplitude is related to the branching ratio of
the decay $\phi\to\mu^+\mu^-$ by the following formula:
\begin{equation}
Q=\frac{3\sqrt{B(\phi\to e^+e^-)B(\phi\to\mu^+\mu^-)}}{\alpha},
\label{Bmm}
\end{equation}
where $\alpha$ is the fine structure constant. 
From this relation the product is \\
$\sqrt{B(\phi\to e^+e^-)B(\phi\to\mu^+\mu^-)}
=(3.14\pm0.22\pm0.14)\cdot 10^{-4}$.
Using the
table value  $B(\phi\to e^+e^-)=(2.99\pm0.08)\cdot 10^{-4}$
\cite{PDG}, we obtain
$B(\phi\to\mu^+\mu^-)=(3.30\pm0.45\pm0.32)\cdot 10^{-4}$.

\section{Conclusion}
\normalsize 
In this work the energy dependence of 
the cross section of the process $e^+e^-\to \mu^+\mu^-$ 
in the vicinity of the $\phi$ resonance was studied.
The interference pattern determined by the 
decay $\phi\to \mu^+\mu^-$ was measured, giving
the product of the leptonic branching ratios of $\phi$ meson \\
$\sqrt{B(\phi\to e^+e^-)B(\phi\to\mu^+\mu^-)}
=(3.14\pm0.22\pm0.14)\cdot 10^{-4}$. 
Assuming $\mu-e$-universality one can compare this value with 
the table branching ratio  
 $B(\phi\to e^+e^-)=(2.99\pm0.08)\cdot 10^{-4}$ \cite{PDG}
They are in good agreement. Not using $\mu-e$-universality 
the branching ratio 
 $B(\phi\to\mu^+\mu^-)=(3.30\pm0.45\pm0.32)\cdot 10^{-4}$ was obtained.
It is close to one standard deviation from the
 table value  $B(\phi\to\mu^+\mu^-)=(2.5\pm0.4)\cdot 10^{-4}$ \cite{PDG}.

\section{Acknowledgement}
\normalsize
This work is supported in part by Russian Fund for basic 
researches (grant 96-15-96327) and 
STP "Integration" (No.274).

\end{document}